\documentclass[a4]{article}
\usepackage{graphicx}
\usepackage{amsmath}
\usepackage{amssymb}

\begin{document}
\title{New Photodetection Method Using Unbalanced Sidebands for Squeezed Quantum Noise in Gravitational Wave Interferometer}
\author{K. Somiya\\
\textit{Advanced Material Sciences, Univ. of Tokyo, Bunkyo-ku, Tokyo 113-0033, Japan}}

\maketitle

\begin{abstract}
Homodyne detection is one of the ways to circumvent the standard quantum limit for a gravitational wave detector. In this paper it will be shown that the same quantum-non-demolition effect using homodyne detection can be realized by heterodyne detection with unbalanced RF sidebands. Furthermore, a broadband quantum-non-demolition readout scheme can also be realized by the unbalanced sideband detection.
\end{abstract}

\section{Introduction}
Currently several large-scale LASER interferometric gravitational wave detectors have been developed. There are many kinds of noise sources that limit a detector's sensitivity. All of these noise sources other than the quantum noise may be reduced by technical advances in the near future. The LASER beams used in these interferometers have coherent light which satisfies the least uncertainty relation between the number of photon and the phase of the light, which represents the particle and wave nature of light, respectively. Fluctuations in the photon number produce radiation pressure noise on the test masses and fluctuations in phase produces shot noise. The shot noise can be reduced by increasing the LASER power while the radiation pressure noise is enhanced. Consequently there exists a sensitivity limit which is called the standard quantum limit (SQL). Actually there are several ways to circumvent this SQL, called squeezing, by unbalancing the coherent state. In this paper, we introduce a new detection method to realize a sensitivity better than the SQL which does not require any additional equipment for the detection scheme other than a planned control scheme of a detuned interferometer which will be discussed later in this paper.

\section{Standard Quantum Limit}
Fluctuations in the photon number $\Delta n$ and the phase of light $\Delta\phi$ are related to each other by Heisenberg's uncertainty principle. A coherent state is a quantum state in which the product of these two uncertainties is minimized as $\Delta n\cdot\Delta\phi = 1/2$. The electric field of a LASER beam can be represented roughly by the following equation.
\begin{eqnarray}
E_{\mathrm{in}}&=&\sqrt{N+\Delta n}\ e^{-i(\Omega t+\Delta\phi)}\nonumber\\
&\simeq&\left(1+\frac{\Delta n}{2N}+i\Delta\phi\right)\sqrt{N}e^{-i\Omega t}\label{eq:b1b2}
\end{eqnarray}
Here, $N$ is the average photon number and $\Omega$ is the light frequency. From statistical theory, $\Delta n=\sqrt{N}$, thus, $\Delta\phi=1/2\sqrt{N}$. Thus the fluctuations of the real-part and imaginary-part of eq. (\ref{eq:b1b2}) are equal, indicating that the light is in a coherent state.

\begin{figure}[h]
\begin{center}
 \includegraphics*[height=8cm]{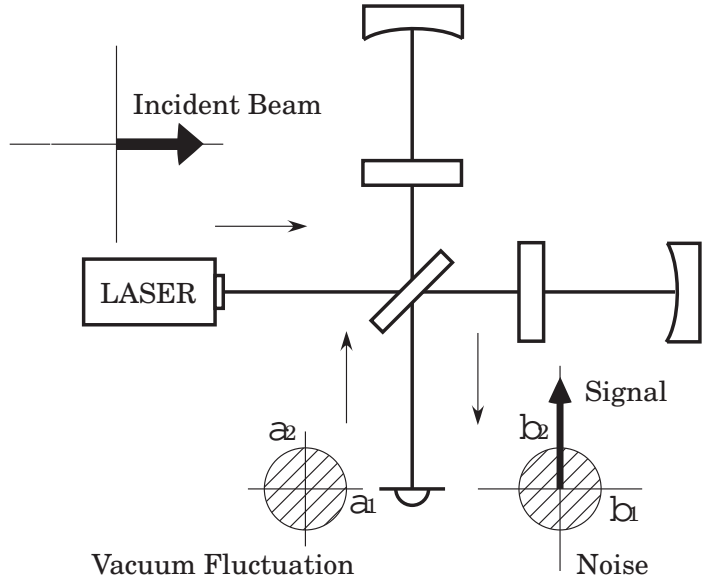}
 \caption{The quantum noise of incident coherent light results from the vacuum fluctuation. The coordinate $a_1$ is the source of photon number fluctuation and $a_2$ is the source of phase fluctuation. Each of these coordinates consists of a pair of annihilation and creation operators.\label{fig:vacuum}}
\end{center}
\end{figure}

The quantum fluctuation in each observation frequency $\omega$ consists of a pair of sideband fields whose frequencies are $\Omega\pm\omega$. Each sideband is excited by the influence of the annihilation and creation operators. When the sidebands are added, the fluctuation is that of photon number; when they are subtracted, it is the phase fluctuation. As two excited electric fields with different frequency are not independent it is natural to use the two-photon mode description\cite{2photon}, in which a pair of photons are considered simultaneously. There are two combinations of the photon excitation:
\begin{eqnarray}
a_1&=&\frac{a_++a_-^\dagger}{\sqrt{2}}\\
a_2&=&\frac{a_+-a_-^\dagger}{\sqrt{2i}}
\end{eqnarray}
Here, $a$ and $a^\dagger$ are the annihilation and creation operators, respectively, the suffix $\pm$ indicates frequency of upper and lower operators. Let us define $a_1$ in the same quadrature as the incident light and $a_2$ in the other quadrature. Then $a_1$ and $a_2$ mean the source of photon number and light phase fluctuations respectively, which are the quadrature coordinates for the source of optical quantum noise, called vacuum fluctuations\cite{vacuum}\cite{vacuum2}. The vacuum fluctuations can be injected from any transmissive optic such as beam splitter or lossy mirrors. The anti-symmetric port of a Michelson interferometer for a gravitational wave detector is kept dark to optimize the signal-to-noise ratio against the shot noise; all the light but the gravitational wave signal is reflected to the symmetric port. These two ports are called the dark port and the bright port, respectively. For simplicity, we assume there is no leaking carrier light to the dark port, which means a perfect-contrast interferometer. Hence force the vacuum fluctuations coming from the direction of the LASER are reflected to the bright port; only the vacuum from the direction of the signal detection port can contribute quantum noise (Fig.\ref{fig:vacuum}). Moreover for simplicity we assume there is no optical loss in the interferometer; the reflectivity for the end mirrors in the arm cavities are set to unity.

The output field of the dark port consists of differential components including a gravitational wave signal and fluctuations in the light which are derived from a coupling between the incident beam and the vacuum fluctuations coming from the dark port. From eq.(\ref{eq:b1b2}) one may think that the quadrature coordinates of the output field $b_1$ and $b_2$ are balanced equally, but actually theie amplitudes differ because of a coupling from photon number fluctuations to phase noise, that is called radiation pressure noise.

According to Ref.\cite{KLMTV}, we define $\kappa(\omega)$ as a coupling parameter between $a_1$ and the radiation pressure noise,
\begin{eqnarray}
\left (
\begin{array}{@{\,}c@{\,}}
b_1\\
b_2
\end{array}\right )
=\left (
\begin{array}{@{\,}cc@{\,}}
1&0\\
-\kappa(\omega)&1
\end{array}\right )
\left (
\begin{array}{@{\,}c@{\,}}
a_1\\
a_2
\end{array}\right ) e^{2i\lambda}
\end{eqnarray}
\begin{eqnarray}
\kappa(\omega) =\frac{(I_0/I_{\mathrm{SQL}})2\gamma^4}{\omega^2(\gamma^2+\omega^2)}.\label{kappa}
\end{eqnarray}
Here, the interferometric configuration is assumed to be a recombined Michelson interferometer with Fabry-Perot resonator in the arms, which is the design of TAMA and Initial-LIGO for example. The phase shift $2\lambda$ is gained by the signal sideband during a round trip in the arm cavity, $\gamma$ is the Fabry-Perot cavity pole frequency, $I_0$ is the light power at the beam splitter, and $I_{\mathrm{SQL}}=mL^2\gamma^4/4\Omega$ is the light power necessary to reach the SQL sensitivity at $\omega =\gamma$, when $m$ is the mass of the mirror and $L$ is the arm cavity length.
\begin{figure}[h]
\begin{center}
 \includegraphics*[height=6cm]{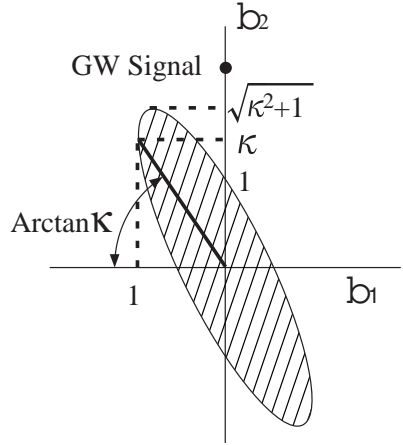}
 \caption{The output quantum noise is squeezed by the effect of the radiation pressure. The quantum noise level is defined as the ratio between the differential signal and the projection of this ellipse to the $b_2$ axis.\label{fig:ellipse}}
\end{center}
\end{figure}

As a result, the quantum output field becomes a squeezed ellipse as shown in Fig.\ref{fig:ellipse}. This is called ponderomotive squeezing. The ellipse is rotated by $(1/2)\mathrm{Arccot}(\kappa/2)$. A local oscillator or RF sidebands used to measure the amount of the signal. The signal is proportional to the amplitude of the local oscillator. After the photodetection, the squeezed vacuum combines with the local oscillator to generate the quantum noise which is also proportional to the amplitude of the local oscillator. The amplitude of the input vacuum fields $a_1$ and $a_2$ is one. The noise level is defined as the ratio between the projection of the quantum noise ellipse to the $b_2$ axis and the signal appearing on that axis. The signal is proportional to the amplitude of the incident classical field $\sqrt{\kappa}$, so the sensitivity is limited by
\begin{eqnarray}
h_\mathrm{n}=h_\mathrm{SQL}\cdot\sqrt{\frac{\kappa^2+1}{2\kappa}},\label{eq:hnlimit}
\end{eqnarray}
where $h_\mathrm{SQL}\equiv\sqrt{8\hbar/m\omega^2L^2}$ is the SQL sensitivity, which occurs when $\kappa=1$, or $I_0=I_\mathrm{SQL}$ and $\omega=\gamma$. One may think this ponderomotive squeezing effect makes it possible to overcome the SQL but the limit is defined including this squeezing effect\cite{BraginskySQL}.

\section{Conventional Detection and Homodyne Detection}
For the last few decades, several theoretical efforts to overcome the SQL have been made\cite{Input}\cite{Homodyne}\cite{Speedmeter}\cite{Alessandra}\cite{Kawamura}, called Quantum Non-Demolition (QND)\cite{QM}. Among these, homodyne detection is a way to cancel the radiation pressure noise effect. As mentioned above, conventional photodetection can only measure a signal and noise on the $b_1$ or $b_2$ axes, but homodyne detection can measure any phase allowing us to readout where the noise is less - the thin part of the ellipse.
 
In conventional detection we use RF sidebands for a modulation-demodulation scheme.
\begin{eqnarray}
\mathrm{Signal}&:&2iD\sin{\psi(t)}\nonumber\\
\mathrm{Sidebands}&:&-4i\alpha\cos{\omega_\mathrm{m}t}\sin{\eta}\nonumber\\
\mathrm{Vacuum}&:&b_1\cos{(\xi(t)+\phi_1)}+b_2i\sin{(\xi(t)+\phi_2)}\nonumber
\end{eqnarray}

Here the common factor $e^{-i\Omega t}$ is omitted for simplicity (just as is shown in the phasor diagram). $D$ is the classical amplitude of the carrier, $\psi(t)$ is the phase shift by gravitational waves, $\alpha$ is the amplitude of the sidebands, $\eta$ is the phase shift caused by an asymmetry which is necessary to leak the sidebands to the dark port, $b_1=1$ and $b_2=\sqrt{\kappa^2+1}$ are the amplitudes of the common and the differential mode operators of a squeezed vacuum, $\xi$ is an ellipse parameter which represents random noise, and the phase shifts $\phi_1$ and $\phi_2$ are given as follows.
\begin{eqnarray}
\theta&=&\frac{1}{2}\mathrm{Arccot}\frac{\kappa}{2}\nonumber\\
B_1^2&=&(\kappa\sin{\theta}-\cos{\theta})^2+\sin{\theta}^2\nonumber\\
B_2^2&=&(\kappa\cos{\theta}-\sin{\theta})^2+\cos{\theta}^2\nonumber\\
\phi_1&=&\arctan{[{B_2\sin{\theta}}/B_1\cos{\theta}]}\\
\phi_2&=&\arctan{[{B_1\sin{\theta}}/B_2\cos{\theta}]}
\end{eqnarray}

The signal, sidebands and vacuum components shown above are summed and squared by the photodetector.
\begin{eqnarray}
&&\left | \mathrm{Signal}+\mathrm{Sidebands}+\mathrm{Vacuum} \right |^2\nonumber\\
&&=4b_2D\sin{\psi(t)}\sin{(\xi(t)+\phi_2)}+b_1^2\cos^2{(\xi(t)+\phi_1)}+b_2^2\sin^2{(\xi(t)+\phi_2)}+4D^2\sin^2{\psi(t)}\nonumber\\
&&\ \ -8\alpha b_2\underline{\cos{\omega_\mathrm{m} t}}\sin{(\xi(t)+\phi_2)}\sin{\eta}-16D\alpha\sin{\psi(t)}\underline{\cos{\omega_\mathrm{m} t}}\sin{\eta}+16\alpha^2\cos^2{\omega_\mathrm{m} t}\sin^2{\eta}\label{conventional}
\end{eqnarray}
Extracting the underlined components at a frequency of $\omega_\mathrm{m}$ by multiplying by $\cos{\omega_\mathrm{m}t}$, a process called demodulation, we obtain the signal-to-noise ratio as follows. The random noise parameter $\xi(t)$ is time-averaged.

\begin{eqnarray}
-8\alpha\frac{\sqrt{\kappa}}{2}\frac{h}{h_\mathrm{SQL}}-2\sqrt{2}\alpha\sqrt{\kappa^2+1}\label{eq:noise}
\end{eqnarray}
Here we define a signal produced by strain $h$ according to
\begin{eqnarray}
D\sin{\psi(t)}=\frac{\sqrt{\kappa}}{2}\frac{h}{h_\mathrm{SQL}}.
\end{eqnarray}
And then the noise spectral density reads
\begin{eqnarray}
S_h=h_n^2=\frac{h_\mathrm{SQL}^2}{2}\left (\kappa+\frac{1}{\kappa}\right ),
\end{eqnarray}
which indicates that we cannot overcome the SQL with conventional detection.

In homodyne detection, the local oscillators are combined with the signal light both with and without a $\pi$ phase shift. This can be realized by placing a beam splitter at the signal extraction port and injecting the incident light and the signal from opposite sides of the beam splitter. The squeezed vacuum injected from the output beam splitter is cancelled by this process and the local oscillator can be considered as a classical light.
\begin{eqnarray}
\mathrm{LO}&:&\beta\cos{\zeta}+\beta\mathrm{i}\sin{\zeta}\nonumber,
\end{eqnarray}
where $\beta$ is the amplitude of the local oscillator and $\zeta$ is a homodyne phase, i.e. the phase of the local oscillator relative to the phase of the carrier light. Subtraction of the two interfered beams after square-law detection gives the following equation.
\begin{eqnarray}
&&\frac{1}{2}\left \{\left | \mathrm{Vac}+\mathrm{Sig}+\mathrm{LO} \right |^2-\left | \mathrm{Vac}+\mathrm{Sig}-\mathrm{LO} \right |^2\right\}\nonumber\\
&=&2\beta\sin{\zeta}\left [\sqrt{\frac{1+(\kappa-\cot{\zeta})^2}{2}}+2D\sin{\psi(t)}\right ]\label{homodyne}
\end{eqnarray}
The noise spectral density reads
\begin{eqnarray}
S_h=\frac{h_\mathrm{SQL}^2}{2}\frac{1+(\kappa-\cot{\zeta})^2}{\kappa},\label{eq:Sh}
\end{eqnarray}
which  has a minimum value when the homodyne phase is $\zeta=\mathrm{arccot}{\kappa}$. This result has been introduced in previous papers\cite{KLMTV}\cite{Alessandra}. We can see that the radiation pressure effect disappears for a particular frequency band (Fig.\ref{fig:spectrum}).

\section{Unbalanced Sideband Detection}
Now the author will show that the same performance achieved by homodyne detection can also be achieved by using only a single sideband - without any additional optics such as the output beam splitter.
\begin{eqnarray}
\mathrm{Single}\ \mathrm{Sideband}\ :\ -2i\alpha e^{i\omega_\mathrm{m}t}\sin{\eta}\nonumber
\end{eqnarray}

Eq.(\ref{conventional}) will have a different solution as follows.
\begin{eqnarray}
&&\left | \mathrm{Signal}+\mathrm{Single}\ \mathrm{Sideband}+\mathrm{Vacuum}\right |^2\nonumber\\
&&=4b_2D\sin{\psi(t)}\sin{(\xi(t)+\phi_2)}+4D^2\sin^2{\psi(t)}-4\alpha b_2\underline{\cos{\omega_\mathrm{m} t}}\sin{(\xi(t)+\phi_2)}\sin{\eta}\nonumber\\
&&\ \ -8\alpha D\sin{\psi(t)}\underline{\cos{\omega_\mathrm{m} t}}\sin{\eta}+4\alpha b_1\underline{\sin{\omega_\mathrm{m}t}}\cos{(\xi(t)+\phi_1)}\sin{\eta}\nonumber\\
&&\ \ +b_1^2\cos^2{(\xi(t)+\phi_1)}+b_2^2\sin^2{(\xi(t)+\phi_2)}+4\alpha^2\sin^2{\eta}\label{USB}
\end{eqnarray}
After the same demodulation process as the conventional detection scheme, i.e. by multiplying by $\sin{(\omega_\mathrm{m}-\zeta')}$, the output is
\begin{eqnarray}
2\alpha\sin{\zeta'}\left [\sqrt{\frac{1+(\kappa-\cot{\zeta'})^2}{2}}+2D\sin{\psi(t)}\right ]
\label{eq:output}
\end{eqnarray}
which is the same equation as eq.(\ref{homodyne}) except $\alpha\rightarrow\beta$ and $\zeta\rightarrow\zeta'$. Notice that $\zeta'$ is not a homodyne phase but a heterodyne demodulation phase, the phase shift between the modulation sidebands and the RF demodulation signal. Conventionally the demodulation phase is set to be $\pi/2$ to obtain a differential signal, but considering the ponderomotive squeezing effect the phase should be set to be $\zeta'=\mathrm{arccot}{\kappa}$ to minimize the quantum noise at the desired observing frequency band.

\begin{figure}[h]
\begin{center}
 \includegraphics*[height=6cm]{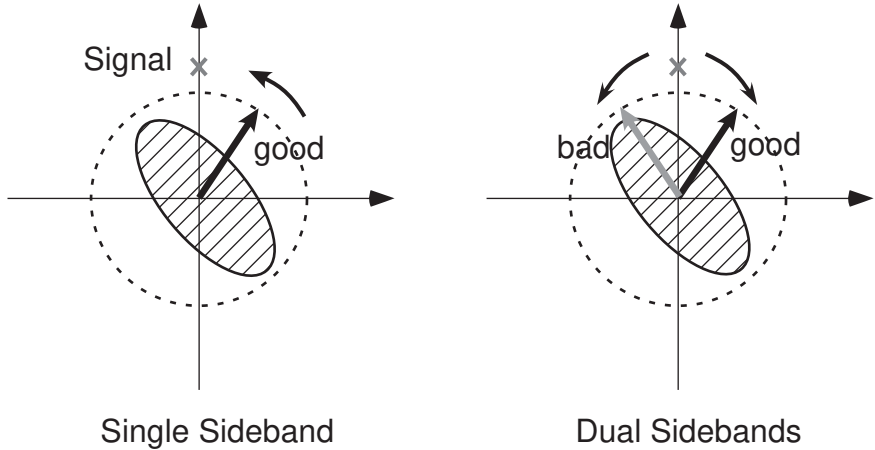}
 \caption{Signal detection can be done at any phase for single sideband detection, but for conventional dual sidebands detection the other sideband will produce a worse signal-noise ratio.\label{fig:SingleSB}}
\end{center}
\end{figure}

One may wonder why this single sideband detection can beat the SQL even though it is impossible with conventional dual sidebands detection. Figure \ref{fig:SingleSB} explains the reason clearly. With conventional detection using dual sidebands, even if one wants to optimize the demodulation phase for one sideband, the other sideband phase is anti-optimal and generates increased noise. As a result the total noise is equivalent to the case when the sidebands stay on the differential axis.

This QND detection can also be realized with dual sidebands if there is an unbalance between the amplitude of the upper and lower sidebands. It can be derived that the noise spectrum in that case is the same as Eq.(\ref{eq:Sh}) when the demodulation phase meets the following condition.
\begin{eqnarray}
\zeta'=\mathrm{arccot}\frac{\alpha}{\Delta\alpha}\kappa
\end{eqnarray}
Here, $\alpha$ and $\Delta\alpha$ are the average and the difference of the amplitude of the upper and the lower sidebands. When the unbalance factor $\Delta\alpha$ is small, the output (\ref{eq:output}) is small, although the signal-to-noise ratio is unaffected.

There are several ways to produce unbalance of the sidebands, one of which is to make only one sideband resonate in a cavity placed at the dark port. Advanced-LIGO will employ what is called a detuned resonant-sideband-extraction (RSE) system, realized with a signal recycling cavity detuned from the resonant condition. To control the detuned cavity, it is planned to make only the upper or lower sideband resonant, hence, the carrier is detuned from resonance\cite{LIGOII}. The transmittance of the RSE cavity for the resonant sideband is larger than the non-resonant sideband, producing the necessary sideband unbalance.

One outstanding advantage for this detection technique is the possibility of broadband QND detection by combining several outputs each optimized for a different frequency. On the phasor diagram the single sideband rotates by its modulation frequency; 12 megahertz in Initial-LIGO, for example. Here the demodulation is a kind of the stroboscopic measurement\cite{QM}, clipping the moving signal-to-noise ratio at some periodic phase and producing a spectrum for that phase. There is no quantum demolition if the measured \textit{photocurrent} is split to several demodulator circuits, each of which has a different demodulation phase. There is, on the other hand, quantum demolition if the output \textit{beam} is split to several photodetectors, because another vacuum fluctuation would be injected from the optical components used for the division. In other words, if several optical local oscillators with different homodyne phases were to be used to try to beat the SQL at several frequency bands, the decrease in optical power would increase the shot noise level, but if different-phase electrical local oscillators are used there is no corresponding increase in shot noise.

\begin{figure}[h]
\begin{center}
 \includegraphics*[height=6cm]{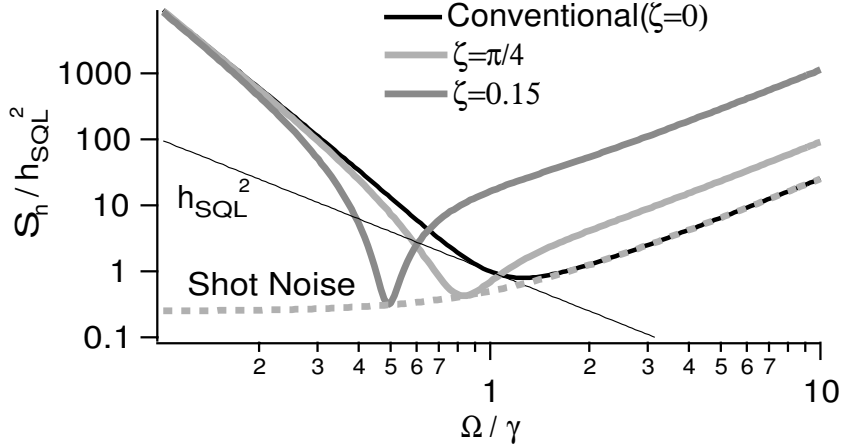}
 \caption{Using homodyne detection, it is possible to overcome the SQL for a limited frequency band. The center frequency can be changed by adjusting the homodyne phase. The minimum value of each spectrum is limited only by shot noise, i.e. quantum noise without radiation pressure noise. \label{fig:spectrum}}
\end{center}
\end{figure}

Ideally, by measuring all of the signal with different demodulation phases, one can obtain just the same quantum noise spectrum as is shown for the dotted line in Fig.\ref{fig:spectrum}. This indicates there is no radiation pressure noise - only shot noise. We can call this method the broadband QND detection.

\section{Conclusions and Discussions}In this paper, the author shows that the single sideband detection, or more generally unbalanced sideband detection, can realize the same improvement in the quantum noise level achieved by homodyne detection. Moreover, broadband QND can be realized by this new photodetection method using multi-channel demodulation. Applying the multi-channel technique to a conventional non-detuned interferometer, we can remove the radiation pressure noise completely.

It should be mentioned that there is an extra shot noise contribution which can decrease signal-to-noise ratio by nearly 20\%. It is neglected in this paper for simplicity. In the conventional RF scheme and the unbalanced sideband detection, the vacuum fluctuations at a frequency of $\Omega\pm 2\omega_\mathrm{m}$ combine with the RF sidebands and generate this extra shot noise, which is called nonstationary shot noise\cite{Niebauer}\cite{Strain}\cite{Mio}. For the conventional interferometric configuration, it is reported in these papers that the nonstationary shot noise can be reduced by using a square-wave form for modulation and/or demodulation instead of a sinusoidal wave. Further study is needed for the detuned interferometer\cite{Nergis}.

The author has mentioned that one advantage of this method is that the unbalanced sideband condition is naturally realized using only the planned control scheme for a detuned RSE configuration. Indeed, the quantum noise spectrum for the detuned RSE will be quite different from a conventional interferometer's noise spectrum\cite{Alessandra}. For the detuned RSE, ponderomotively squeezed vacuum is reflected by the RSE mirror and returns to the dark port with some phase shift, so it can be a sort of input squeezing conditions; this squeezing make it possible to circumvent the SQL at some frequencies, and consequently the quantum noise curve can show two dips. This phenomenon is called the optical spring. Both dips can be moved by changing the homodyne phase. 

\begin{figure}[h]
\begin{center}
 \includegraphics*[height=8cm]{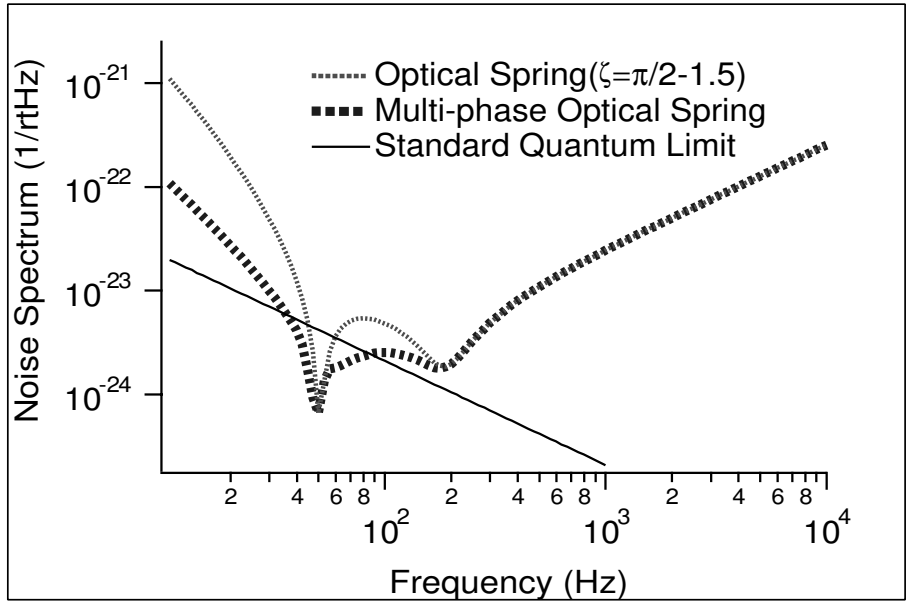}
 \caption{Broadband QND. The laser power is 500W, the finesse of the arm cavities is about 2000, the reflectivity of the RSE mirror is 0.98, and the detuning phase is 0.03 degrees from the RSE condition. The homodyne phase or the demodulation phase for the small dotted line is set to $\pi/2-1.5$, and that for the large dotted line is chosen from $\pi/2$, $\pi/2-1.2$, and $\pi/2-1.5$ to be optimized for each frequency; this can be called a "multi-phase optical spring." \label{fig:BroadBand}}
\end{center}
\end{figure}

A quantum noise spectrum for detuned RSE with multi-channel single sideband detection is shown in Fig.\ref{fig:BroadBand}. Here we use three signal ports with different demodulation phases for example. One can see the noise spectrum is improved, expanding from a narrow band to a broadband. Further broadband detection can be realized by sweeping the detuning phase, a technique that will be developed in near future.

A practical multi-phase detection method for a high-power detuned RSE should be studied more. The optical spring can generate instabilities between the arm cavities and the transmissive RSE mirror\cite{Alessandra2}. Such instabilities can be cured, for example in the case with a simple homodyne detection, by damping the optical spring with a feed-back control system. The condition will be more complicated with a multi-phase detection scheme or with sweeping, so further studies are needed.

\section{acknowledgments}
The author wishes to thank Professor S.Kawamura, Professor N.Mio, and O.Miyakawa for giving the author a chance to think about this method. And also the author wishes to thank members of Advanced-Interferometer-Configuration Committee from LSC (LIGO Science Collaboration) for many helpful comments, especially P.Beyersdorf who helped edit this paper. This research is supported by a Grant-in-Aid for Creative Basic Research of the Ministry of Education, Science, Sports and Culture.

\bibliographystyle{junsrt}

\end{document}